# Classification of Cellular Automata Rules Based on Their Properties

Pabitra Pal Choudhury [1], Sudhakar Sahoo[2], Sarif Hasssan[3], Satrajit Basu[4], Dibyendu Ghosh[5], Debarun Kar[6], Abhishek Ghosh[7], Avijit Ghosh[8] , Amal K. Ghosh[9]
[1,2,3] Applied Statisctics Unit,
Indian Statistical Institute,
Kolkata- 700108, India
[1,2]Email: pabitrapalchoudhury@gmail.com & sudhakar.sahoo@gmail.com ,
[4,5,6,7,8,9] Netaji Subhash Engineering College,
Technocity, Garia, Kolkata- 700152, India
[3,4,5,6,7]Email: satrajit.87@gmail.com, babu.education@gmail.com, kar_debarun@yahoo.co.in, ghosh_abi4u@yahoo.co.in, avijit2all@yahoo.com, amal_k_ghosh@rediffmail.com

**Abstract.** This paper presents a classification of Cellular Automata rules based on its properties at the nth iteration. Elaborate computer program has been designed to get the nth iteration for arbitrary 1-D or 2-D CA rules. Studies indicate that the figures at some particular iteration might be helpful for some specific application. The hardware circuit implementation can be done using opto-electronic components [1-7].

## 1 Introduction

The wide range of applications of Cellular Automata (CA)[8-16] used in various fields like Computer Science, Physics, Biology, Chemistry, Mathematics, Social Science and Engineering etc. has raised curiosity to examine the structure of the set of CA rules in a more systematic way. A CA is an idealized parallel processing machine consisting of a number of cells containing some cell values called states together with an updating rule. A cell value is updated based on this updating rule, which involves the cell value as well as other cell values in a particular neighborhood. The CA is completely defined with the help of following 5 parameters.

1. Number of *States*
2. Size of *Neighborhood*
3. *Length* of the CA
4. Guiding *Rules* (Uniform or hybrid) and
5. Number of *times* the evolution takes

These 5-parameters affect the number of CA rules. General formulae for computing Number of CA rules by changing different parameters as shown in Table-1.

**Table-1** : Number of CA rules

| | |
|---|---|
| Possible Number of Uniform CA rules | $State^{State^{Neighborhood}}$ |
| Possible Number of Hybrid CA rules | $\left(State^{State^{Neighborhood}}\right)^{length\ of\ the\ CA}$ |
| Possible Number of Hybrid CA rules changing over different time steps throughout the evolution | $\left(\left(State^{State^{Neighborhood}}\right)^{length\ of\ the\ CA}\right)^{time}$ |
| Possible Number of Hybrid CA rules changing over different time steps throughout the evolution and each cell can also change its neighborhood at each iteration | $\left(S_1^{S_1^{N_1}} \times S_2^{S_2^{N_2}} \ldots \times S_i^{S_i^{N_i}} \ldots \times S_n^{S_n^{N_n}}\right)^{time}$ |

The number of uniform CA rules increases astronomically with the increase in neighborhoods as shown in Table-2 below.

**Table-2**: Number of uniform CA rules

| Neighborhood | 4 | 5 | 6 | 7 | 8 |
|---|---|---|---|---|---|
| Uniform CA rules | $2^{16}$ | $2^{32}$ | $2^{64}$ | $2^{128}$ | $2^{256}$ |
| ~ | $6.10^4$ | $4.10^9$ | $10^{19}$ | $10^{38}$ | $10^{77}$ |

So it is quite impossible to explore this astronomical number of rules even with the fastest computer available today unless the computing paradigm changes dramatically. So one can raise an important question as: *How to find the exact CA rule(s), which will model a particular application?* One *Solution to this problem is – Classification.*

Wolfram [9] classified 256 possible 1-D, 3-neighborhood 2-states CA rules into four classes as Type 1, Type 2, Type 3 and Type 4 based on their complexities as demonstrated in the space-time diagram. Many others have also used the same classification technique for the analysis of different CA rules. In this paper, we are motivated to explore different tools to analyze systematically the set of CA rules so that the theory developed during our research work can be used for some suitable applications.

With this background the current research work has been undertaken as outlined next. Here, we have tried to classify all the CA rules on the basis of their properties and we propose a new method of classification different from the way Wolfram [9] classified all the CA rules based on its space-time diagram. In this paper our objective is to design an automatic classifier based on the properties of the snapshot obtained by the CA rules in the $n^{th}$ evolution. This is same as evaluating each individual contestant (CA rule) in a competition based on its present performance rather than it's past.

At present the optical computation is more advantageous because of its faster operational speed with huge volume of data. There is a fundamental difference between optics and electronics – electron being Fermions are mutually interactive while photon being Boson such interactions are to be employed in an indirect manner.

Again due to this non-interaction characteristics of photon, communication of stream of data in 1-D and 2-D images is possible. Thus our implementations in 1-D and 2-D CA can be fruitfully done using opto-electronics devices.

The organization of the paper is as follows. Section 2 discusses the review of earlier works. Classification of 512 linear rules based on the pattern observed in the $n^{th}$ iteration is given section 3. It can be found in section 4 that all the 512 linear CA rules can be classified based on the discrete and continuous nature of the generated patterns obtained in the $n^{th}$ iteration. Experimental evidence for both the classifications given in section 3 and section 4 are mentioned in section 5. Finally a conclusion is drawn in section 6.

## 2 Review of the earlier works

In 2-D Nine Neighborhood CA the next state of a particular cell is affected by the current state of itself and eight cells in its nearest neighborhood also referred as Moore neighborhood as shown in Figure-1. Such dependencies are accounted by various rules. For the sake of simplicity, in this section we take into consideration only the linear rules, i.e. the rules, which can be realized by EX-OR operation only. A specific rule convention that is adopted here is as follows [8]:

Considering the nearest neighbor concept of 2-D Cellular Automata (CA) there are 9 variables to be taken under consideration as shown:

| | | |
|---|---|---|
| 64 | 128 | 256 |
| 32 | 1 | 2 |
| 16 | 8 | 4 |

**Fig.1**: Rule convention chosen for 2-D CA

The middle cell marked '**1**' is the cell under consideration. In 2-D Cellular Automata, the state of the cell under consideration depends upon its own state and the state of its neighboring cells. Now each of the cells can be taken as a variable and thus for 2-D cellular Automata there are 9 variable to be considered. The number of linear rules can be realized by EX-OR operation only. The number of such rules generated by a combination of these 9-variables is ${}^{9}C_0 + {}^{9}C_1 + . . . + {}^{9}C_9 =512$ which includes rules characterizing no dependency.

Now, these 512 linear rules have been previously classified by taking into account the number of cells under consideration. The grouping has been Group-N for N=1, 2…. 9, includes the rules that refer to the dependency of current cell on the N neighboring cells amongst top, bottom, left, right, top-left, top-right, bottom-left, bottom-right and itself. Thus group 1 includes 1, 2, 4, 8, 16, 32, 64, 128, and 256. Group 2 includes 3, 5, 6, 9, 10, 12, 17, 18, 20, 24, 33, 34, 36, 40, 48, 65, 66, 68, 72, 80, 96, 129, 130, 132, 136, 144, 160, 192, 257, 258, 260, 264, 272, 288, 320 and 384.

Similarly rules belonging to other groups have been obtained. It can be noted that number of 1's present in the binary sequence of a rule is same as its group number.

## 3 A New insight separate from S. Wolfram's Work

There is a significant paradigm shift between our approach and the Wolfram's approach. In the space-time diagram of 1-D CA after finite number of iterations, Wolfram took into account the patterns generated for each iteration and visualized it in a 2-D plane. This when extended to the 2-D domain results in the stacking of each of the plane figures generated in successive iterations one on top of the other to give rise to a three-dimensional figure, which shows a nested structure. Now, Wolfram classified the one-dimensional CA rules based on their complexity. For that he took into account the behavior of the rules in all the steps till the $n^{th}$ iteration starting from the seed value.

But, we will try and classify the CA rules based on the behavior of the rule in the $n^{th}$ iteration. Next section gives an example based on this type of classification.

### 3.1 Classification of 512 linear CA rules based on the patterns obtained in the $n^{th}$ iteration

Now our goal was to try and classify 512 rules into different groups. First we studied the patterns being generated for a given seed for all the rules iterating it for a fixed number of times. We first considered an 80X80 matrix with the Null boundaries. For our experiment we chose a single initial seed and its location was fixed at (40, 40). We then applied each 2-D CA linear rule separately on the initial seed till n= $(2^k-1)$ number of iterations, where k is an integer. The patterns generated in this way produces different plane figures such as a point, a straight line, a triangle, a quadrilateral, a pentagon or a hexagon. We classified the rules on the basis of these generated patterns. The entire classification for n=15 is shown in Table 4.

**Table**-4

| Rules | Pattern |
|---|---|
| 1, 2, 4, 8, 16, 32, 64, 128, 256 | * |
| 3, 5, 6, 9, 10, 12, 17, 18, 20, 24, 33, 34, 35, 36, 40, 48, 65, 66, 68, 69, 72, 80, 96, 112, 129, 130, 132, 136, 137, 144, 152,160,190, 192, 257, 258, 260, 261, 262, 264, 272, 273, 288, 320, 384, 448 | **************** |
| 7, 11, 13, 14, 19, 21, 22, 25, 26, 28, 29, 30, 37, 38, 39, 41, 42, 43, 44, 49, 50, 51, 52, 56, 60, 67, 70, 73, 76, 77, 81, 82, 84, 85, 88, 92, 93, 97, 98, 99, 100, 101, 104, 116, 117, 120, 124, 125, 131, 133, 134, 138, 140, 141, 145, 146, 148, 149, 153, 156, 157, 161, 162, 163, 164, 165, 168, 176, 180, 193, 194, 196, 197, 200, 201, 208, 224, 240, 259, 263, 265, 266, 267, 268, 270, 274, 276, 277, 278, 279, 280, 281, 284, 285, 286, 287, 289, 290, 291, 292, 293, 294, 295, 296, 297, 304, 308, 309, 310, 311, 321, 322, 324, 325, 326, 327, 328, 329, 336, 337, 352, 360, 368, 369, 385, 386, 388, 390, 392, 393, 400, 401, 416, 432, 433, 449, 450, 452, 453, 454, 455, 456, 457, 464, 465, 480, 488, 489, 496, 497 | 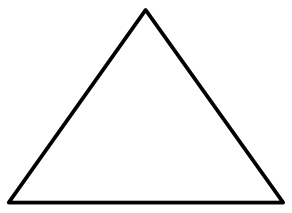 |
| 15, 23, 27, 31, 45, 46, 47, 53, 54, 55, 57, 58, 59, 61, 62, 63, 71, 74, 75, 83, 102, 103, 105, 106, 107, 108, 109, 113, 114, 115, 135, 139, 147, 150, 166, 167, 169, 170, 171, 176, 177, 178, 179, 181, 195, 194, 204, 205, 212, 213, 216, 217, 220, 221, 225, 226, 227, 228, 229, 230, 231, 232, 233, 244, 245, 249, 252, 253, 269, 271, 275, 298, 298, 299, 300, 301, 302, 305, 306, 307, 312, 313, 316, 317, 323, 332, 333, 334, 335, 340, 341, 342, 343, 344, 345, 348, 349, 350, 351, 353, 354, 355, 356, 357, 358, 359, 361, 364, 365, 366, 367, 372, 373, 374, 375, 376, 377, 380, 381, 382, 383, 387, 389, 391, 396, 397, 398, 399, 404, 405, 406, 407, 408, 409, 412, 413, 414, 415, 417, 418, 419, 420, 421, 422, 423, 424, 425, 436, 437, 438, 439, 440, 444,445, 446, 447, 451, 460, 461, 462, 463, 468, 469, 470, 471, 472, 473, 476, 477, 478, 479,481, 482, 483, 484, 485, 486, 487, 492, 493, 494, 495, 500, 501, 502, 503, 504, 505, 508, 509, 510, 511 | 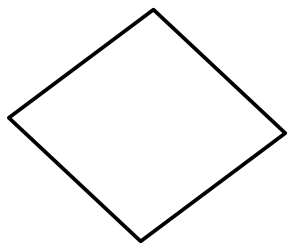 |
| 78, 79, 86, 87, 89, 90, 91, 94, 95, 110, 111, 118, 119, 121, 122, 123, 126, 127, 142, 143, 151, 154, 155, 158, 159, 172, 173, 174, 182, 183, 186, 187, 188, 189, 190, 191, 198, 199, 202, 203, 206, 207, 209, 210, 211, 214, 215, 218, 219, 222, 223, 234, 235, 236, 241, 242, 243, 246, 247, 250, 251, 254, 255, 282, 283, 251, 303, 314, 315, 318, 319, 330, 331, 338, 339, 346, 347, 362, 363, 370, 371, 378, 379, 394, 395, 402, 403,410,411,426, 427, 428, 429, 430, 431, 434, 435, 458, 459, 466, 467, 474, 475, 490, 491, 498, 499, 506, 507 | 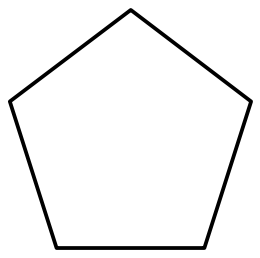 |
| 236, 237, 238, 239, 441,442, 443 | 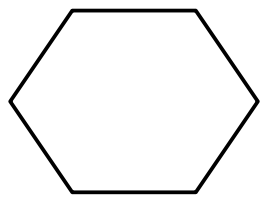 |

It may be noted from [14], that CA rules may also be classified on the basis of their capacities in producing multiple similar figures.

In the above table each pattern is generated for a fixed seed by the application of the CA rules on each cell in the Matrix for n iterations. Let us consider for example how the pattern for rule 21 is generated. For rule 21, the cells under consideration are highlighted as shown in Fig2.

| | | |
|---|---|---|
| 64 | 128 | 256 |
| 32 | 1 | 2 |
| 16 | 8 | 4 |

**Fig.2**: Cells considered for rule 21

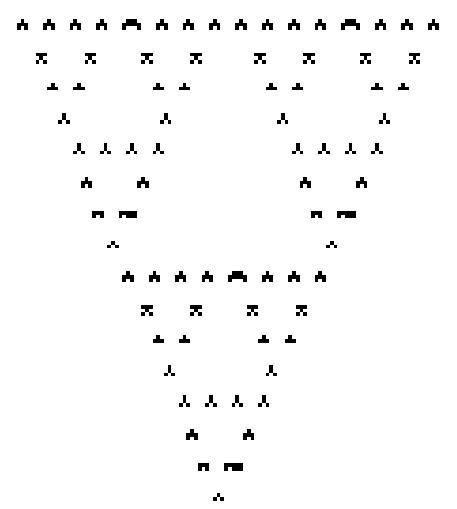

**Fig 3**: Generated pattern for Rule-21 at 15$^{th}$ iteration.

On applying this rule on an initial seed at (40, 40) yields the above figure 3 at the 15$^{th}$ iteration.

## 4 Classification of 512 linear CA rules based on the discrete and continuous nature of the generated patterns obtained in the n$^{th}$ iteration

Now, the patterns generated for each of the 512 linear 2-D CA rules can be grouped as discrete or continuous based on the nature of the pattern generated. A pattern is termed *continuous* if in the space-time diagram there exists a continuous path from one vertex to another. Else, it is termed as *discrete*. Based on this, the previous classification, as obtained in [14] can be further sub-classified as *Discrete* or *Continuous.*

The classification for n= 15 is given in Table 5.

**Table**-5

| |
|---|
| **Group 1**<br>Discrete: 1, 2, 4, 8, 16, 32, 64, 128, 256 |
| **Group 2**<br>Discrete: 18, 20, 34, 66, 68, 72, 80, 132, 136, 144, 192, 257, 260, 264, 272, 288,320<br>Continuous: 3, 5, 6, 9, 10, 12, 17, 24, 33, 40, 48, 65, 129, 257, 258, 384 |
| **Group 3**<br>Discrete: 21, 22, 28, 35, 38, 42, 50, 52, 69, 76, 84, 88, 100, 104, 112, 137, 140, 148, 152, 162, 196, 200, 208, 232, 262, 268, 273, 276, 280, 290, 292, 296, 304, 322, 324, 328, 336, 352, 388, 392, 400, 448<br>Continuous: 7, 11, 13, 14, 19, 25, 26, 37, 41, 44, 49, 56, 67, 73, 74, 81, 82, 97, 98, 104, 131, 133, 134, 138, 145, 146, 161, 164, 168, 176, 193, 194, 224, 259, 265, 266, 289, 292, 321, 385, 386, 416 |
| **Group 4**<br>Discrete: 29, 30, 39, 43, 46, 51, 53, 54, 58, 60, 77, 83, 85, 87, 90, 92, 99, 101, 106, 108, 116, 120, 141, 149, 153, 154, 156, 163, 166, 172, 178, 180, 184, 197, 201, 204, 212, 216, 226, 228, 240, 263, 269, 270, 277, 278, 281, 284, 291, 293, 294, 298, 300, 305, 308, 312, 323, 325, 326, 329, 330, 332, 353, 354, 356, 360, 368, 389, 390, 393, 396, 401, 404, 408, 418, 420, 424, 432, 449, 450, 452, 456, 464, 480.<br>Continuous: 15, 23, 27, 45, 57, 71, 75, 78, 86, 89, 102, 105, 114, 135, 139, 142, 147, 150, 165, 170, 177, 195, 202, 209, 210, 225, 267, 279, 297, 306, 387, 394,402, 417 |
| **Group 5**<br>Discrete: 31, 47, 55, 59, 61, 62, 79, 91, 103, 108, 109, 110, 115, 117, 118, 121, 122, 124, 143, 151, 155, 157, 158, 171, 173, 174, 181, 185, 186, 188, 199, 203, 205, 206, 211, 213, 214, 217, 218, 220, 227, 229, 230, 233, 234, 236, 241, 242, 244, 248, 271, 279, 283, 285, 286, 295, 301, 302, 303, 307, 309, 310, 313, 314, 316, 327, 331, 333, 334, 339, 341, 342, 345, 346, 348, 355, 357, 358, 361, 362, 364, 369, 370, 372, 376, 391, 395, 397, 398, 403, 405, 406, 409, 410, 412, 419, 421, 422, 425, 426, 428, 433, 434, 436, 440, 453, 454, 457, 458, 460, 465, 466, 468, 472, 481, 482, 484, 488, 496,<br>Continuous: 107, 167, 179, 451 |
| **Group 6**<br>Discrete: 63, 95, 119, 125, 126, 159, 182, 183, 187, 190, 207, 215, 221, 222, 231, 237, 245, 246, 249, 252, 287, 311, 317, 318, 335, 343, 347, 349, 350, 359, 365, 366, 373, 374, 377, 378, 380, 399, 407, 413, 414, 423, 429, 430, 435, 437, 438, 444, 455, 459, 461, 462, 467, 469, 470, 473, 474, 476, 483, 485, 486, 489, 490, 492, 497, 498, 500, 504.<br>Continuous: 111, 123, 175, 187, 219, 231, 235, 238, 243, 250, 315, 365, 371, 411, 442, 427, 435, 441 |
| **Group 7**<br>Discrete: 127, 191, 223, 247, 253, 254, 319, 351, 367, 375, 379, 381, 382, 415, 431, 439, 445, 446, 463, 471, 475, 477, 478, 487, 491, 493, 494, 499, 501, 502, 505, 506, 508<br>Continuous: 239, 251, 443. |
| **Group 8**<br>Discrete: 255, 383, 447, 479, 495, 503, 507, 509, 510. |
| **Group 9**<br>Discrete: 511 |

# 5 Experimental Results

```
****************
* * * * * * * *
**  **  **  **
*   *   *   *
****    ****
* *     * *
**      **
*       *
********
* * * *
**  **
*   *
****
* *
**
*
```

**Fig 4**: pattern for Rule=25 at 15$^{th}$ iteration

```
* * * * * * * * * * * * * * * *
* * * * * * * * * * * * * * * *
* * * * * * * * * * * * * * * *
* * * * * * * * * * * * * * * *
* * * * * * * * * * * * * * * *
* * * * * * * * * * * * * * * *
* * * * * * * * * * * * * * * *
* * * * * * * * * * * * * * * *
* * * * * * * * * * * * * * * *
* * * * * * * * * * * * * * * *
* * * * * * * * * * * * * * * *
* * * * * * * * * * * * * * * *
* * * * * * * * * * * * * * * *
* * * * * * * * * * * * * * * *
* * * * * * * * * * * * * * * *
* * * * * * * * * * * * * * * *
```

**Fig 5:** pattern for rule 54 at 15$^{th}$ iteration.

For Rule 25, in Table-4 this rule is grouped under rules producing patterns having triangular shape [Fig 4]. In Table-5 this rule is grouped under Group 3 as the cells under consideration for Rule 25 are 16, 8, and 1. In Table-5 it is also grouped under continuous, as the pattern generated in the 15$^{th}$ iteration is a continuous one.

For Rule 54, in Table-4 this rule is grouped under rules producing patterns similar to quadrilaterals [Fig 5]. In Table-5 this rule is grouped under Group 4 as the cells under consideration for Rule 54 are 32, 16, 4 and 2. In Table-5 it is also grouped under Discrete, as the pattern generated in the $15^{th}$ iteration is a discrete one.

The points to be noted here is that the patterns are termed continuous and discrete based on the graphical definition of vertices where one vertex can be reached from another or not respectively. Again, the patterns being generated can be considered as Fractals only if the patterns are in the continuous domain as defined by the mathematicians. On the other hand, the researchers view discrete patterns differently. Some researchers impose their own definition for considering them as fractals, while others do not at all consider them as fractals.

## 6 Conclusions

This paper presents a classification of two-dimensional Cellular Automata linear rules based on its properties at the $n^{th}$ iteration. Our studies indicate one potential research area to be pursued for the successful application of a set of CA rules in a particular domain. Further for the purpose of massively parallel information processing, we have adopted the Optical processing scenarios.